\documentclass[preprint,twocolumn]{revtex4-1}

\usepackage{amsmath}
\usepackage{latexsym}
\usepackage{graphicx}
\usepackage{xcolor}
\usepackage{subfig}
\usepackage[linktocpage,colorlinks]{hyperref}
\usepackage{color}

\allowdisplaybreaks[3]

\def\[{\left [}
\def\]{\right ]}
\def\({\left (}
\def\){\right )}

\def\r2{\sqrt{2}}

\newcommand{\beq}{\begin{equation}}
\newcommand{\eeq}{\end{equation}}

\begin{document}
\title{$B$ decays to two pseudoscalars and a generalized $\Delta I = \frac{1}{2}$ rule}
\preprint{UCSD/PTH 14-1}
\author{Benjam\'in Grinstein}
\email{bgrinstein@ucsd.edu}
\affiliation{Department of Physics, University of California, San Diego, La Jolla, CA 92093 USA}
\author{David Pirtskhalava}\email{david.pirtskhalava@sns.it}
\affiliation{Scuola Normale Superiore, Piazza dei Cavalieri 7, 56126 Pisa, Italy}
\author{David C. Stone}
\email{dcstone@physics.ucsd.edu}
\affiliation{Department of Physics, University of California, San Diego, La Jolla, CA 92093 USA}
\author{Patipan Uttayarat}
\email{uttayapn@ucmail.uc.edu}
\affiliation{Department of Physics, University of Cincinnati, Cincinnati, OH 45220 USA}
\affiliation{Department of Physics, Srinakharinwirot University, Wattana, Bangkok 10110 Thailand}

\begin{abstract}
We perform an isospin analysis of $B$ decays to two pseudoscalars. The
analysis extracts appropriate CKM and short distance loop factors to allow for
comparison of non-perturbative QCD effects in the reduced matrix elements of the amplitudes. In decays where penguin diagrams compete with tree-level diagrams we find that
the reduced matrix elements of the penguin diagrams, which are singlets or doublets under isospin, are significantly enhanced compared with the
triplet and fourplet contributions of the weak Hamiltonian. This similarity to the $\Delta I =\frac{1}{2}$ rule in $K \to \pi \pi$ decays suggests that, more
generally, processes mediated by Hamiltonians in lower-dimensional isospin representations see enhancement over higher-dimensional ones in QCD. 
\end{abstract}

\maketitle
\section{Introduction}
One of the longstanding puzzles in flavor physics is the $\Delta
I=1/2$ rule. An isospin-$\tfrac12$ neutral  kaon may decay into 
 two pions  in either an isospin-0 or isospin-2  ($s$-wave) state
 with amplitude $A_0$ or $A_2$, respectively. Empirically,
 \beq\label{eq:deltaIhalf}
\frac{\text{Re}\,A_0}{\text{Re}\,A_2}=22.5~.
\eeq
The $\Delta I=1/2$ rule is the statement that the amplitude $A_0$,
mediated by the part of the weak Hamiltonian that transforms as an
$I=1/2$ tensor, is much larger than $A_2$, mediated by the larger
$I=3/2$ tensor. 

There is no satisfactory understanding of this rule.  In
Refs.~\cite{Bardeen:1986vp,Bardeen:1986uz,Bardeen:1986vz} and, more recently, Ref.~\cite{Buras:2014maa} the rule was
investigated in chiral perturbation theory, in the large $N_c$
limit. However, it was argued in Ref.~\cite{Chivukula:1986du} that for QCD, $N_c =3$ is not large enough for this limit to be useful.
More recent studies using Monte Carlo simulations of QCD in the
lattice have addressed the $\Delta I=1/2$ rule
\cite{Blum:2011ng}; a very recent study on the lattice of the validity of the vacuum insertion approximation was done in \cite{Carrasco:2013jda}. 
The ratio in \eqref{eq:deltaIhalf} is still twice as large as any values obtained on the lattice with unphysical quark masses, but
it is expected that simulations at physical quark masses will reproduce the empirically observed ratio and shed light on the origin of the enhancement
\cite{Boyle:2012ys}. This begs the question--
does this enhancement occur in systems other than the $K \to \pi\pi$ system?

There is evidence that answers this question in the affirmative. Identifying any patterns of enhancements will give new insights into the long distance
dynamics of QCD. For example, the $SU(3)$ analysis of $D\to KK,\pi\pi$
decays reveals a similar enhancement. In that system, the $D^0\to
K^+K^-$ and $D^0\to\pi^+\pi^-$ amplitudes may be written as \cite{Quigg:1979ic}
\begin{align}
\mathcal{A}(D^0\to K^+K^-) &= (2T+E-S)\Sigma \nonumber \\
& +\tfrac12(3T+2G+F-E)\Delta \nonumber \\
\mathcal{A}(D^0\to \pi^+\pi^-)&=- (2T+E-S)\Sigma \nonumber \\
& +\tfrac12(3T+2G+F-E)\Delta \nonumber
\end{align}
where
$\Sigma\equiv\tfrac12(V^*_{cs}V^{\phantom{*}}_{us}-V^*_{cd}V^{\phantom{*}}_{ud})$
and
$\Delta\equiv\tfrac12(V^*_{cs}V^{\phantom{*}}_{us}+V^*_{cd}V^{\phantom{*}}_{ud})$. 
$S$, $E$ and $F$ are the invariant matrix elements between a $D$ meson and
a meson pair in an octet of the $\mathbf{\bar 6}$, {\bf 15} and {\bf
  3} components of the weak Hamiltonian, respectively,  $G$ of the
{\bf 3} to a singlet pair and $T$ of the {\bf 15} to a meson pair in
the {\bf 27}. Note that $\Sigma\approx \lambda=\sin\theta_C$, while
$|\Delta|\sim\lambda^5$, so that
$|\Delta|/\Sigma\sim10^{-3}$. Neglecting $\Delta$ one would have
$\Gamma(D^0\to K^+K^-)=\Gamma(D^0\to \pi^+\pi^-)$ in the $SU(3)$
limit. Experimentally $\Gamma(D^0\to K^+K^-)/\Gamma(D^0\to
\pi^+\pi^-)\approx 3$ requires both the $\Sigma$ and $\Delta$ terms in
the amplitude to contribute with similar strengths. Barring accidental
cancellations this means that the matrix elements $G$ and $F$ are
significantly enhanced. Since $\Delta$ has a large phase, 
significant CP-violation in these decays was predicted \cite{Golden:1989qx} and recently confirmed by experiment
\cite{Aaltonen:2011se,Collaboration:2012qw,Aaij:2013bla}.  

If $SU(3)$-breaking effects are included, the ratio $\Gamma(D^0\to K^+K^-)/\Gamma(D^0\to
\pi^+\pi^-)\approx 3$ can be attained with only a  ``mild'' enhancement of $F$ and $G$ relative to the other reduced matrix elements of about an order of magnitude
\cite{Pirtskhalava:2011va,Bhattacharya:2012ah,Feldmann:2012js,Brod:2012ud,Cheng:2012xb,Hiller:2012xm}. The enhancement in $F$ and $G$ is similar
to that of the $\Delta I=1/2$ rule in that it appears in matrix
elements of the smallest $SU(3)$-representation of the Hamiltonian. In
this case, the dominant contributions are from the {\bf 3} Hamiltonian (as opposed to the $\mathbf{\bar 6}$ and {\bf
  15}), whereas for the $\Delta I=1/2$ rule the dominant piece is from
the $I=1/2$ Hamiltonian (as opposed to the $I=3/2$ piece).

In this work we investigate the possibility of similar enhancements
in $B$ decays. We will show that an isospin analysis of $B\to K\pi$
decays and CP-asymmetries shows a marked enhancement of amplitudes
mediated by the weak Hamiltonian in the lowest isospin representation. 
An analysis of $B\to \pi\pi$ decays shows that, although there is
little enhancement of doublet versus fourplet amplitudes, the matrix
elements of penguin contributions (which are purely $\Delta I=1/2$) are
still enhanced to produce the observed data. 
Both these analyses
support the general rule that amplitudes mediated by the piece of the
weak Hamiltonian in the smallest representation of the symmetry group
are enhanced. 

It should go without saying that we have no dynamical explanation of
the enhancement. This comes as no surprise, since the very $\Delta I
=1/2$ rule has resisted explanation for more than a half century. But
we hope that insights provided by this new, generalized rule may
eventually lead to a global understanding of these enhancements.

\section{Isospin analysis}
The strong interactions, to a good approximation, obey isospin
symmetry.  In hadronic spectra and decays
isospin violating effects are no larger than a few per cent.  We study
the amplitudes for the decay of $B$-mesons to two light scalar mesons
using isospin symmetry, under which kaons and $B$-mesons transform as doublets
and pions as a triplet. The possible
two-body final states are easily classified according to their
transformation properties under isospin. We also need 
the transformation properties of the effective Hamiltonian responsible for the
weak decay.  The effective Hamiltonian  is given in terms of four-quark operators, whose transformation
properties are readily determined.  

\subsection{$B\to K\pi$}\label{bkpi}
The effective Hamiltonian density for the $\Delta B=-1$, $\Delta S=-1$ decays, to leading order in the Fermi constant $G_F$, can be written as \cite{Buras:1998raa,Buchalla:1995vs}
\begin{align}
\label{hamiltonianBKpi}
\mathcal{H} = \frac{G_F}{\sqrt{2}} & \left[ \lambda_u \(C_1 Q_1 + C_2Q_2\) 
-\lambda_t\sum_{i=3}^{6} C_i Q_i \right].
\end{align}
Here $\lambda_q\equiv V^*_{qb}V_{qs}$ are CKM factors and $C_i$'s are the
Wilson coefficients. The ``tree'' ($Q_{1,2}$) and ``penguin''
($Q_{3-6}$) operators are defined as
\begin{align}
\label{operators}
Q_1  & = \(\bar b_a u_b\)_{V-A} \(\bar u_b s_a\)_{V-A}, \nonumber\\ 
Q_2  & = \(\bar b u\)_{V-A} \(\bar u s\)_{V-A}, \nonumber\\
Q_3  & = \(\bar b s\)_{V-A} \sum_{q=u,d}\(\bar q q\)_{V-A}, \nonumber\\
Q_4  & = \(\bar b_a s_b\)_{V-A} \sum_{q=u,d}\(\bar q_b q_a\)_{V-A}, \nonumber\\
Q_5  & = \(\bar b s\)_{V-A} \sum_{q=u,d}\(\bar q q\)_{V+A}, \nonumber\\
Q_6  & = \(\bar b_a s_b\)_{V-A} \sum_{q=u,d}\(\bar q_b q_a\)_{V+A} 
\end{align}
where $(\bar qq)_{V\pm A}$ is shorthand for $\bar
q\gamma^\mu(1\pm\gamma^5)q$.  Both the coefficients $C_i$ and the
matrix elements of the operators $Q_i$ depend on an arbitrary
renormalization point $\mu$ but their combination in the Hamiltonian,
Eq.~\eqref{hamiltonianBKpi}, is $\mu$-independent. QCD-penguins
arising from $u$ and $c$ quark loops combine into terms precisely of
the form of top-quark penguins, since
$\lambda_c+\lambda_u=-\lambda_t$.  We have also neglected electroweak
penguins (EWP), operators $Q_{7-10}$ in Ref.~ \cite{Buras:1998raa}. These
introduce new isospin triplets into the Hamiltonian with a
$\lambda_t$ coefficient, suppressed relative to the top-penguins by
$\alpha/\alpha_s$. We have ignored EWP contributions out of pragmatism: were we to
include their effects in our fits the number of unknown
matrix elements would exceed the number of measured data. But our
pragmatism is informed: the coefficients of EWP in the effective
Hamiltonian are suppressed relative to QCD penguins roughly by a
factor of $\alpha/\alpha_s$, or about 7\% if evaluated at $\mu = M_Z$
and smaller at $m_b$.  As will become evident, the approximation is supported by the very good
fit of the model to {\it both} $B\to K\pi$ and $B\to \pi\pi$
processes.

	\begin{centering}
\begin{table*}
	\centering
	\begin{tabular}{c@{\hspace{.3cm}}  |@{\hspace{.3cm}} c@{\hspace{.3cm}} c@{\hspace{.3cm}}c@{\hspace{.3cm}}c }
		\hline
		\hline
		Mode & $\mathcal{B}$ ($10^{-6}$) & $A_{CP}$ & $C_f$ & $S_f$\\
		\hline
		$B^+ \to K^+ \pi^0$ & $12.9\pm0.5$  & $\phantom{-}0.037 \pm 0.021$ & --& -- \\
		$B^+ \to K^0 \pi^+$ & $23.8 \pm 0.7$  & $-0.014 \pm 0.019$ & -- & -- \\
		$B^0_d \to K^0  \pi^0$ & $\phantom{1}9.9\pm 0.5$  & -- & $0.00 \pm 0.13 $ & $0.58 \pm 0.17$  \\
		$B^0_d \to K^+ \pi^-$ & $19.6\pm0.5$  & $-0.087 \pm 0.008$ & -- & --  \\
		\hline
	\end{tabular}
	\caption{Data available in $B\to K\pi$ decays~\cite{Beringer:1900zz}. The $C$ and $S$ parameters are measured for decays into the final CP eigenstate, $B^0_d\to K^0_s\pi_0$. The 
amplitude for $B^0_d\to K^0\pi^0$ on the other hand is given as $\mathcal{A}(B^0_d\to K^0\pi^0)=\sqrt{2}~ \mathcal{A}(B^0_d\to K^0_s\pi^0).$}
	\label{tab:bkpitab}
\end{table*}
\end{centering}

As far as the group theory analysis of rates and CP asymmetries is
concerned, different four-quark operators contributing to the
Hamiltonian can be distinguished solely by their isospin quantum
numbers and CKM factors. The Hamiltonian can therefore be compactly
written in terms of the isospin representations in the following way:
\beq
H=V^*_{ub}V_{us}~\(\mathbf{1}+[\mathbf{3}]^1_{~1}\)+\frac{\alpha_s}{8\pi}~V^*_{tb}V_{ts}~\mathbf{1'}~,
\eeq 
where $\mathbf{1}$ ($\mathbf{1'}$)  denotes
the singlet coming from the tree (penguin) operators, $
[\mathbf{3}]^1_{~1} $ represents the triplet
operator, and $\alpha_s$ the strong coupling constant evaluated at
$M_Z$.  We choose to normalize the singlet penguin operator with an agnostic 
factor of $\alpha_s/(8\pi)$ to make explicit the loop factor
associated with it. This normalization does not affect the results of this paper, but it is a useful choice that, na\"ively, would give reduced
matrix element values of the same order of magnitude for every contribution.
%
We introduce shorthand for  the reduced matrix elements, as follows: 
\begin{align}
\label{amplitudes}
& \langle \mathbf{\bar 2|1|}B\rangle \equiv P_b, \quad \langle \mathbf{\bar 2|\mathbf{1'}|}B\rangle \equiv P_a, \nonumber \\
& \langle \mathbf{\bar 2|3|}B\rangle \equiv T, \quad  \langle \mathbf{\bar 4|3|}B\rangle \equiv S~. 
\end{align}
While we cannot compute  $P_a$, $P_b$, $S$ and $T$ from first
principles, we can determine them by fitting to experimental
measurements of decay rates and CP asymmetries.

\begin{figure}[h]
	\centering
	\includegraphics[width=.45\textwidth]{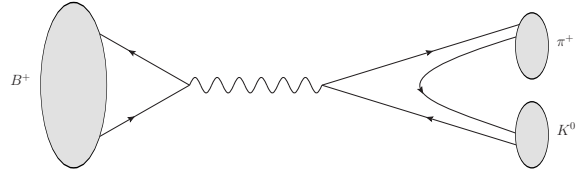}
	\caption{Leading order diagram contributing to the $B^+ \to K^0 \pi^+$ process.}
	\label{fig:annihilationDiagram}
\end{figure}

In terms of the reduced matrix elements in Eq.~\eqref{amplitudes}, the
isospin decomposition of the decay amplitudes is 
\begin{align}
\label{eq:BKpi}
\mathcal{A}(B^+\to K^+\pi^0) 
&  =
V^{\ast}_{ub}V^{\phantom{\ast}}_{us}~\frac{1}{\sqrt{2}}\left(P_b+T+2S\right)
\nonumber\\
&+ \,\frac{\alpha_s}{8\pi}~V^{\ast}_{tb}V^{\phantom{\ast}}_{ts}~\frac{P_a}{\sqrt{2}}~,\nonumber\\
\mathcal{A}(B^+\to K^0\pi^+) 
& = V^{\ast}_{ub}V^{\phantom{\ast}}_{us}~\left(P_b+T-S\right)\nonumber \\
&+ \,\frac{\alpha_s}{8\pi}~V^{\ast}_{tb}V^{\phantom{\ast}}_{ts}~P_a~,\nonumber\\
\mathcal{A}(B^0\to K^0\pi^0) 
&  =
V^{\ast}_{ub}V^{\phantom{\ast}}_{us}~\frac{1}{\sqrt{2}}~\left(-P_b+T+2S\right)\nonumber \\
&- \,\frac{\alpha_s}{8\pi}~V^{\ast}_{tb}V^{\phantom{\ast}}_{ts}~\frac{P_a}{\sqrt{2}},\nonumber\\
\mathcal{A}(B^0\to K^+\pi^-) 
& = V^{\ast}_{ub}V^{\phantom{\ast}}_{us}~\left(P_b-T+S\right) \nonumber\\
&+ \,\frac{\alpha_s}{8\pi}~V^{\ast}_{tb}V^{\phantom{\ast}}_{ts}~P_a ~.
\end{align}
There is a contribution proportional to
$V^{\ast}_{ub}V^{\phantom{\ast}}_{us}$ to the amplitude
$\mathcal{A}\(B^+\to K^0\pi^+\)$.  The only contribution to this
process stems from the annihilation diagram, shown in
Fig.~\ref{fig:annihilationDiagram}.  There is extensive literature on
annihilation diagram suppression with respect to $W$-emission diagrams
\cite{Chau:1990ay,Gronau:1994bn}. To evaluate this expectation, denote
the matrix element associated with the annihilation diagram
by $M\equiv P_b+T-S$ and let $|M|=x |P_a|$ so that $x$ measures the
relative importance of  annihilation  in comparison to the
top-loop penguin. The value of $x$ for which the
annihilation and penguin contributions to $B^+\to K^0\pi^+$ are of the
same order can be estimated as 
\beq
\label{eq:xzero}
x=\frac{\alpha_s}{8\pi}~\bigg|\frac{V^{\ast}_{tb}V^{\phantom{\ast}}_{ts}}{V^{\ast}_{ub}V^{\phantom{\ast}}_{us}}\bigg |\simeq 0.24~.
\eeq

\begin{figure*}
	\centering
	\subfloat[]{\label{fig:PaPb}\includegraphics[width=0.45\textwidth]{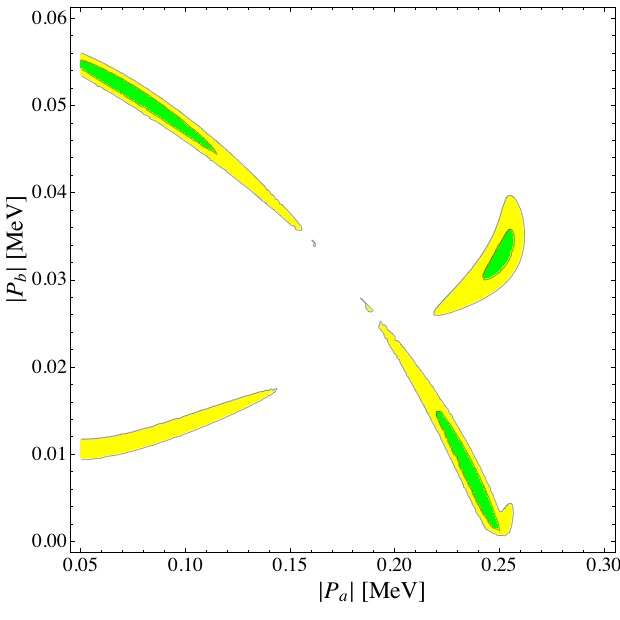}}
	\subfloat[]{\label{fig:PaPb}\includegraphics[width=0.45\textwidth]{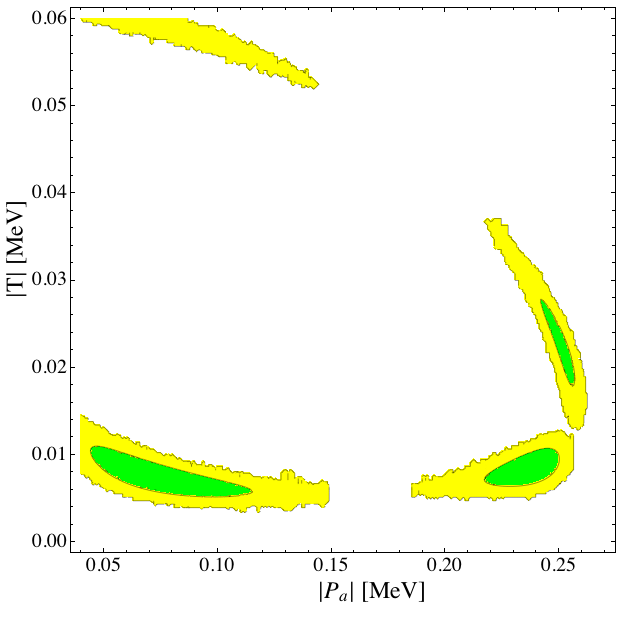}}
	\caption{Fit to data of the reduced matrix elements for $B\to K\pi$. The
          figures show 
        the 68\% (green) and 95\% (yellow) CL regions in the $|P_a|$
        vs $|P_b|$ and $|P_a|$
        vs $|T|$  planes. 
      The raggedness of the contours is an artifact of the numerical computation.}
	\label{fig:KpiCLplots}
\end{figure*}

\subsection*{Results of the fit}
The available decay data for $B\to K\pi$ are collected in Table~\ref{tab:bkpitab}; the observables are defined in
Appendix~\ref{sec:sonvention}.  Performing a $\chi^2$ fit of matrix
elements in Eq.~\eqref{eq:BKpi} to the data, we find values for the matrix elements that match the observed data with a 95\% confidence level. 
These minima are illustrated with 68\% and 95\% confidence levels in the $|P_a|$ vs.~$|P_b|$ and $|P_a|$ vs.~$|T|$ planes, respectively, in Fig.~\ref{fig:KpiCLplots}. 
The best fit has $\{|P_a|,\,|P_b|,\,|T|,\,|S|\}\simeq\{0.237,\,7.2\times10^{-3},\,8.4\times10^{-3},\,2.2\times10^{-3}\}$ MeV with a  chi-squared of ${\chi}^2 = 1.70$ for
two degrees of freedom (a common phase in the reduced matrix elements is unobservable). The $\Delta I=0$  contribution to the amplitudes, from the Hamiltonian in the singlet representation, is given  by the quantity 
\beq
a_{\Delta I = 0} = P_b + \frac{\alpha_s}{8\pi}~\frac{V^{\ast}_{tb}V^{\phantom{\ast}}_{ts}}{V^{\ast}_{ub}V^{\phantom{\ast}}_{us}} P_a
\eeq
and the $\Delta I=1$ contribution,  from the triplet Hamiltonian, by
\beq
a_{\Delta I = 1} = \{T+2S,  T-S\}.
\eeq
for $\(B^+\to K^+\pi^0, B^0\to K^0\pi^+\)$ and $\(B^+\to K^0\pi^+, B^0\to K^+\pi^-\)$ respectively.
For the best fit then, we find 
\beq
\label{enh1}
\bigg| \frac{a_{\Delta I =0}}{a_{\Delta I = 1}}\bigg| = \{4.8,~9.9\}
\eeq
which is reminiscent of the $\Delta I = \tfrac{1}{2}$ rule from $K \to \pi\pi$ decays.

A second, slightly higher $\chi^2$-minimum has $\{|P_a|,\,|P_b|,\,|T|,\,|S|\}\simeq\{0.075,\,0.052,\,7.3\times10^{-3},\,2.4\times10^{-3}\}$ MeV with a  chi-squared of ${\chi}^2 = 1.80$ and 
\beq
\label{enh2}
\bigg| \frac{a_{\Delta I =0}}{a_{\Delta I = 1}}\bigg| = \{5.2,~12.6\} ~.
\eeq

Both of these minima have significant enhancement of the {\em penguin} singlet, $P_a$, over the triplet matrix elements, $T$ and $S$. 
In the best fit case, however, the other singlet matrix element, $P_b$, does not show significant enhancement over the triplet matrix elements. 
Consequently, the annihilation diagram contribution is negligible in the best fit ($|M| = 0.013$ MeV or, equivalently, $x = |M/P_a| = 0.055$, to be
compared with Eq. \eqref{eq:xzero}) but provides a larger contribution than that of the penguin diagram in the second best fit (where $|M| = 0.055$ MeV
or, equivalently, $x = 0.732$).

For completeness we note that there are two additional minima corresponding to $\chi^2 = 3.04$ and 4.34. 
These two minima are less favorable, so we ignore them in the rest of our study.

In all but the least favored minimum, there is significant enhancement of $|P_a|$ over the triplet Hamiltonian matrix elements. 
Moreover, the total contribution from the $\Delta I = 0$ Hamiltonian, $a_{\Delta I = 0}$, enjoys an enhancement over the $\Delta I = 1$
contribution, $a_{\Delta I = 1}$.
More precise data will be welcomed to distinguish between these minima, which would also decide the role of the annihilation diagram in these decays.

\begin{figure*}[t]
	\centering
	\subfloat[]{\label{fig:QpiaQpib}\includegraphics[width=0.45\textwidth]{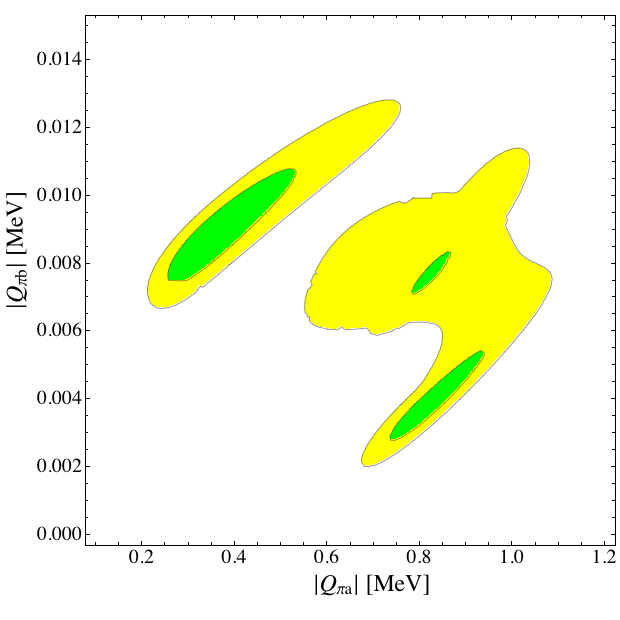}}
	\subfloat[]{\label{fig:QpiaU}\includegraphics[width=0.46\textwidth]{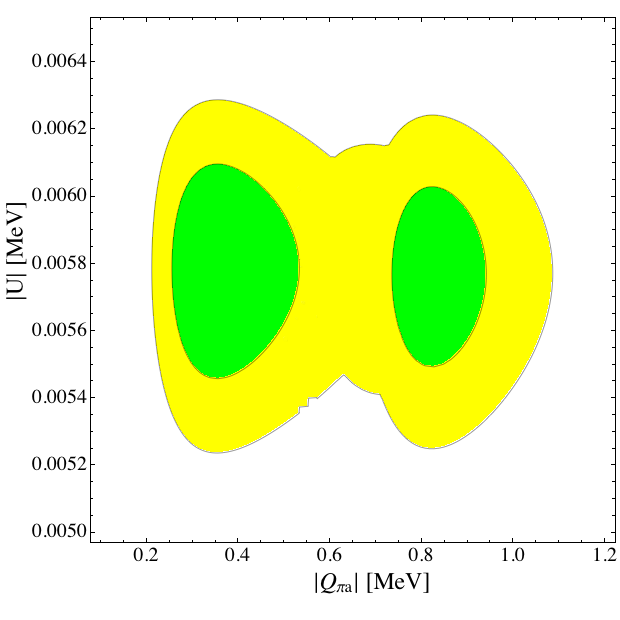}}
	\caption{Fit to data of the reduced matrix elements for
          $B\to\pi\pi$. The
          figures show 
        the 68\% (green) and 95\% (yellow) CL regions in the $|Q_{\pi a}|$
        vs $|Q_{\pi b}|$ and $|Q_{\pi a}|$    vs $|U|$  planes. 
              The raggedness of the contours is an artifact of the numerical computation.}
	\label{fig:pipiCLplots}
\end{figure*}

\subsection{$B \to \pi \pi$}\label{bpipi}
The isospin analysis for $\pi\pi$ final states is analogous to that
for $K$ decays, where the $\Delta I =\tfrac12$ rule was discovered. Operator contributions are of the form in
\eqref{operators}, but for $\Delta S = 0$ processes.  The Hamiltonian decomposes under isospin as $\mathbf{\bar 2\times 2 \times \bar 2
  = \bar 2 + \bar 2 + \bar 4}$ so that
\beq
\label{HDeltas0}
H=V^*_{ub}V_{ud}~\([\mathbf{\bar 2}]^2+[\mathbf{\bar 4}]^{12}_{1}\)+\frac{\alpha_s}{8\pi}~V^*_{tb}V_{ts}~ [\mathbf{\bar 2'}]^2~,
\eeq
The final states transform as $(\mathbf{3} \times \mathbf{3})_S = \mathbf{1} + \mathbf{5}$, 
so the non-vanishing reduced matrix elements are
\beq
\label{matrixpipi}
\langle \mathbf{1|2'|}B\rangle = Q_{\pi a},\; \langle
\mathbf{1|2|}B\rangle = Q_{\pi b},\; \langle
\mathbf{5|4|}B\rangle=U
\eeq
and the decay amplitudes relevant to the processes in Table~\ref{tab:bpipitab} are
\begin{align}\label{amplitudespipi}
	 \mathcal{A}(B^+\to \pi^+\pi^0) &= \sqrt{\frac{3}{2}}~V^{\ast}_{ub}V^{\phantom{\ast}}_{ud}~\,U~,\nonumber\\
         \mathcal{A}(B^0\to \pi^0\pi^0) 
	&  = V^{\ast}_{ub}V^{\phantom{\ast}}_{ud}~\frac{1}{\sqrt{3}}
        \left( Q_{\pi b} - \sqrt{2}\, U \right)\nonumber\\
&+\frac{\alpha_s}{8\pi}~V^{\ast}_{tb}V^{\phantom{\ast}}_{td}~\frac{1}{\sqrt{3}}Q_{\pi a}~,\nonumber\\
	 \mathcal{A}(B^0\to \pi^+\pi^-) 
	&  =
        V^{\ast}_{ub}V^{\phantom{\ast}}_{ud}~\frac{1}{\sqrt{3}}\left(
          \sqrt{2}\,Q_{\pi b} + U \right)\nonumber\\
&+\frac{\alpha_s}{8 \pi}~V^{\ast}_{tb}V^{\phantom{\ast}}_{td}~\sqrt{\frac{2}{3}}\,Q_{\pi a}~.
\end{align}

\subsection*{Results of the fit}

\begin{centering}
\begin{table*}
	\centering
	\begin{tabular}{c@{\hspace{.3cm}}  |@{\hspace{.3cm}} c@{\hspace{.3cm}} c@{\hspace{.3cm}}c@{\hspace{.3cm}}c }
		\hline
		\hline
		Mode & $\mathcal{B}$ ($10^{-6}$)& $A_{CP}$ & $C_f$ & $S_f$\\
		\hline
		$B^+ \to \pi^+ \pi^0$ & $5.5\pm0.4$  & $0.03 \pm 0.04$ &-- & -- \\
		$B^0 \to \pi^0 \pi^0$ & $1.91^\pm0.22$  & -- & $-0.43 \pm 0.24 $ & --  \\
		$B^0 \to \pi^+ \pi^-$ & $5.12 \pm 0.19$  &-- & $-0.38 \pm 0.15$ & $-0.65 \pm 0.07$  \\
		\hline
	\end{tabular}
	\caption{Data available in $B\to \pi\pi$ decays from Ref~\cite{Beringer:1900zz}.}
	\label{tab:bpipitab}
\end{table*}
\end{centering}

The data available in this decay channel are listed in Table ~\ref{tab:bpipitab}.
We perform a $\chi^2$-fit of the model, Eq.~\eqref{amplitudespipi}, to the data. The result of the fit is illustrated with 68\% (green) and 95\% (yellow) CL
regions in the $|Q_{\pi a}|$ vs $|Q_{\pi b}|$ and $|Q_{\pi a}|$ vs
$|U|$ planes, respectively, in
Fig.~\ref{fig:pipiCLplots}. For the best fit to the data we obtain $\{|Q_{\pi
  a}|,~|Q_{\pi
  b}|,~|U|\}\simeq\{0.35,~8.8\times10^{-3},~5.8\times10^{-3}\}$~MeV
with a chi-squared of ${\chi}^2 \simeq 1.39$ for 2 degrees of freedom. 
Two additional regions with a good fit to the data are found, one with $\{|Q_{\pi
  a}|,~|Q_{\pi
  b}|,~|U|\}\simeq\{0.82,~3.9\times10^{-3},~5.8\times10^{-3}\}$~MeV for a chi-squared of ${\chi}^2 \simeq 2.07$  and the other with  $\{|Q_{\pi
  a}|,~|Q_{\pi
  b}|,~|U|\}\simeq\{0.82,~7.7\times10^{-3},~5.8\times10^{-3}\}$~MeV for a chi-squared of ${\chi}^2 \simeq 3.38$.  Since the last minimum is less favorable, we will ignore it.
The contribution to the amplitudes from the Hamiltonian in the doublet
representation is
\beq
a_{\Delta I = 1/2} =
Q_{\pi b} + \frac{\alpha_s}{8\pi} \frac{V^*_{tb}
  V_{td}}{V^*_{ub}V_{ud}} Q_{\pi a} 
\eeq 
and from the fourplet
Hamiltonian 
\beq 
a_{\Delta I = 3/2} = U.  
\eeq 
We find no 
enhancement of the $\Delta I = 1/2$ amplitude with respect to the $\Delta I = 3/2$ amplitude.  To wit, for the best fits (next favorable minimum) we find
\beq
\label{pipienh}
\bigg| \frac{a_{\Delta I = 1/2}}{a_{\Delta I = 3/2}}\bigg| = 1.04\,(1.05).
\eeq
There is little enhancement of the reduced matrix element corresponding to the tree-level doublet Hamiltonian, $Q_{\pi b}$, with respect to the tree-level quadruplet $U$.
However, the large enhancement of the penguin doublet reduced matrix element $Q_{\pi a}$ over $U$ is analogous to that in the $K\to \pi\pi$ decays, which has identical isospin analysis to the
$B\to \pi\pi$ case. That a similar enhancement exists in the $B$
system ---both
in $K\pi$ and $\pi\pi$ final states---  is striking, and cries out for
a dynamical explanation of the role of flavor symmetries in these
enhancements.

\subsection{$B \to K\overline{K}$}\label{bkk}

\begin{centering}
\begin{table*}
	\centering
	\begin{tabular}{c@{\hspace{.3cm}}  |@{\hspace{.3cm}} c@{\hspace{.3cm}} c@{\hspace{.3cm}}c@{\hspace{.3cm}}c }
		\hline
		\hline
		Mode & $\mathcal{B}$ ($10^{-6}$) & $A_{CP}$ & $C_f$ & $S_f$\\
		\hline
		$B^+ \to K^+ \overline{K}^0$ & $1.19 \pm 0.18$  & $0.04 \pm 0.14$ & -- & -- \\
		$B^0 \to K^+ K^-$ & $0.13 \pm 0.05$  & --  & --  & --\\
		$B^0 \to K^0 \overline{K}^0$ & $1.21 \pm 0.16$  & $-0.6\pm0.7$  & $0.0\pm0.4$ & $-0.8 \pm 0.5$\\
		\hline
	\end{tabular}
	\caption{Data available in $B\to K\bar{K}$ decays~\cite{Beringer:1900zz}.}
	\label{tab:bkktab}
\end{table*}
\end{centering}

At leading order, decays of B mesons to kaons proceed via the $\Delta
S=0$ Hamiltonian in \eqref{HDeltas0}.  The $K$ ($\bar{K}$) transforms
as a $\mathbf{2}$ ($\mathbf{\bar{2}}$) under isospin, so the final
states decompose under isospin as $\mathbf{ 2 \times \bar{2} = 1 +
  3}$.

The reduced matrix elements are then $ \langle
\mathbf{1|2^{(\prime)}}|B\rangle$, $\langle
\mathbf{3|2^{(\prime)}}|B\rangle$ and $\langle \mathbf{3|4}|B\rangle$,
giving nine parameters to accommodate the seven data entries listed in Table
\ref{tab:bkktab}. Even with a  measurement of $C$ and $S$   in $B^0\to K^+
K^-$ in hand the matrix elements could not be determined
unambiguously, but with  precise $KK$ data  it may be possible to
distinguish the physical solution from others.

\section{Short distance QCD effects}
How much of the enhancement in the
lower dimensional isospin representation matrix elements can be
attributed to computable short distance QCD effects? 
Comparing the effective Hamiltonian in Eq.~\eqref{hamiltonianBKpi} against  the decay amplitudes in Eq.~\eqref{eq:BKpi}, we see that
\begin{equation}
\label{qcdMEsPa}
\begin{split}
	\frac{\alpha_s}{8\pi}P_a &= \langle K\pi|\sum_{i=3}^6 C_i(m_b)Q_i|B\rangle\\
	& =|C_6(m_b)| \langle\mathbf{2}|\mathbf{1'}|\mathbf{2}\rangle.
\end{split}
\end{equation}
Our analysis cannot yield information about the matrix elements of
each of the operators $Q_{3,\ldots,6}$.  The last step in
\eqref{qcdMEsPa} defines the matrix element of the sum of the
operators, $\langle2|\mathbf{1'}|2\rangle$, after extracting the magnitude of the largest
Wilson coefficient, $|C_6|$. 

Similarly we can define
\begin{equation}
\begin{split}
	P_b & = \langle K\pi|\sum_{i=1,2}C_i(m_b)Q_i|B\rangle = C_-(m_b)\langle\mathbf{2}|\mathbf{1}|\mathbf{2}\rangle,\\
	T &= \langle K\pi|\sum_{i=1,2}C_i(m_b)Q_-|B\rangle = C_-(m_b)\langle\mathbf{2}|\mathbf{3}|\mathbf{2}\rangle,\\
	S &= \langle K\pi|\sum_{i=1,2}C_i(m_b)Q_-|B\rangle = C_-(m_b)\langle\mathbf{4}|\mathbf{3}|\mathbf{2}\rangle,
\end{split}
\end{equation}
where $C_\pm=C_1\pm C_2$ and $Q_\pm=Q_1\pm Q_2$. The $Q_{\pm}$
operators do not have definite isospin. However, for the $B\to\pi\pi$
case the corresponding operator $Q_-$ is pure $\Delta I=1/2$, so using
the $Q_\pm$ basis is natural. Moreover, at 1-loop the operators
$Q_\pm$ do not mix among themselves.  Hence, to estimate the matrix
elements of the ``tree'' operators we have extracted the
coefficient $C_-$. In any case, since $C_\pm$ are of order 1, this introduces
little bias in our analysis.

For our analysis we take the numerical value of Wilson coefficients at NLO in the NDR scheme for $\Lambda^{(5)}_{\bar{MS}}=225$~MeV from table 8 of~\cite{Buras:1998raa}. 
We find that, for matrix elements from our best fit, 
\begin{equation}
\label{finalKpi1}
\begin{aligned}
 	|\langle\mathbf{2}|\mathbf{1'}|\mathbf{2}\rangle| &\approx 0.028 \text{ MeV},\\
	|\langle\mathbf{2}|\mathbf{1}|\mathbf{2}\rangle| &\approx 0.006 \text{ MeV},
\end{aligned}
\;
\begin{aligned}
	|\langle\mathbf{2}|\mathbf{3}|\mathbf{2}\rangle| &\approx 0.007 \text{ MeV},\\
	|\langle\mathbf{4}|\mathbf{3}|\mathbf{2}\rangle| &\approx 0.002 \text{ MeV}.
\end{aligned}
\end{equation}
while for the secondary $\chi^2$ minimum
\begin{equation}
\label{finalKpi2}
\begin{aligned}
 	|\langle\mathbf{2}|\mathbf{1'}|\mathbf{2}\rangle| &\approx 0.009 \text{ MeV},\\
	|\langle\mathbf{2}|\mathbf{1}|\mathbf{2}\rangle| &\approx 0.041 \text{ MeV},
\end{aligned}
\;
\begin{aligned}
	|\langle\mathbf{2}|\mathbf{3}|\mathbf{2}\rangle| &\approx 0.006 \text{ MeV},\\
	|\langle\mathbf{4}|\mathbf{3}|\mathbf{2}\rangle| &\approx 0.002 \text{ MeV}.
\end{aligned}
\end{equation}
The $\Delta I=0$ enhancement for both of these sets of matrix
elements, Eqs.~\eqref{enh1} and~\eqref{enh2}, corresponds to an
enhancement of one or the other singlet matrix element relative to the
largest triplet by a factor of between 4 and 7.  

An analogous analysis can be performed for $B\to\pi\pi$ decays. We define
\begin{align}
\label{qcdMEsQ}
	\frac{\alpha_s}{8\pi}Q_{\pi a} &= \langle \pi\pi|\sum_{i=3}^6 C_i(m_b)Q_i|B\rangle\nonumber\\
	& =|C_6(m_b)| \langle\mathbf{1}|\mathbf{2'}|\mathbf{2}\rangle.\nonumber\\
	Q_{\pi b}  = & \, \langle \pi\pi|\sum_{i=1,2}C_i(m_b)Q_i|B\rangle = C_-(m_b)\langle\mathbf{1}|\mathbf{2}|\mathbf{2}\rangle,\nonumber\\
	U = & \, \langle \pi\pi|C_+(m_b)Q_+|B\rangle = C_+(m_b)\langle\mathbf{5}|\mathbf{4}|\mathbf{2}\rangle,
\end{align}
The matrix element of the operator $Q_+$ can be determined because it is the only ``tree''  contribution to a $\Delta I=3/2$ transition.  We find that, for matrix elements from our best fit, 
\begin{equation}
\label{finalpipi1}
\begin{aligned}
 	|\langle\mathbf{1}|\mathbf{2'}|\mathbf{2}\rangle| \approx 0.040 \text{ MeV}, & \;
	|\langle\mathbf{1}|\mathbf{2}|\mathbf{2}\rangle| \approx 0.007 \text{ MeV}, \\
 	|\langle\mathbf{5}|\mathbf{4}|\mathbf{2}\rangle| \approx & \, 0.006 \text{ MeV},
\end{aligned}
\end{equation}
while for the secondary $\chi^2$ minimum
\begin{equation}
\label{finalpipi2}
\begin{aligned}
 	|\langle\mathbf{1}|\mathbf{2'}|\mathbf{2}\rangle| \approx 0.094 \text{ MeV}, & \;
	|\langle\mathbf{1}|\mathbf{2}|\mathbf{2}\rangle| \approx 0.003 \text{ MeV}, \\
	|\langle\mathbf{5}|\mathbf{4}|\mathbf{2}\rangle| \approx & \, 0.006 \text{ MeV}~.
\end{aligned}
\end{equation}


\section{Discussion and Conclusions}

There is a striking consistency in the reduced matrix element enhancement that persists in the $B$ decay channels studied. As suggested at the end
of Section~\ref{bpipi}, this may be indicative of the importance of flavor symmetries in non-perturbative regimes in QCD, or perhaps in new physics
contributions (note we have only assumed the quark model, CKM parametrization, {\em etc.} of the Standard Model). The enhancement of matrix elements
with effective Hamiltonians in lower-dimensional isospin
representations is only present when penguin diagrams can compete against tree level weak exchanges, which are also the processes where CP violation
is predicted at lowest order. These are the $B \to K\pi$ and $B \to \pi \pi$ channels in this work.

In our estimates for hadronic matrix elements in
Eqs.~\eqref{finalKpi1}, \eqref{finalpipi1} and \eqref{finalpipi2}, but
not \eqref{finalKpi2}, it is  the penguin contributions to 
the lowest isospin change operator  ($\Delta I=0$ for $B\to K\pi$ and $\Delta
I=1/2$ for $B\to\pi\pi$), rather than both penguin and tree contributions, that are enhanced. 
While we cannot select among the fits {\em a priori}, in the best fits for both $B\to K\pi$ and $B\to\pi\pi$  the penguin dominates the total
enhancement, giving a factor of between 4 and 7.  
The precise value of the enhancement is immaterial:  we have made plausible assumptions to remove the short distance QCD effects, but we don't have the means to do this precisely and unambiguously. 
Moreover, the matrix elements $P_a,\ldots, U$ are defined with convenient factors of $\sqrt{2}$ and $\sqrt3$ which further adds to the ambiguity. 
But the enhancement of {\it amplitudes}, Eqs.~\eqref{enh1} (or \eqref{enh2}), is unambiguous. 
Comparable enhancements in the penguin matrix elements for $B\to K\pi$
and $B\to \pi\pi$ lead to a significant amplitude enhancement in $B\to K\pi$ but
very little enhancement in $B\to\pi\pi$,  but only because the latter is CKM-suppressed relative to the former. 


\acknowledgements
DP would like to thank Riccardo Barbieri for valuable discussions. This work was supported in part by the US Department of Energy under contract DE-SC0009919. DP is supported in part by MIUR-FIRB grant RBFR12H1MW. The research of PU has been supported by DOE grant FG02-84-ER40153.

\appendix

\section{Relevant Observables in $B$ Decays}
\label{sec:sonvention}
Here we review the definition of various decay observables employed in our analysis.  We will follow the convention of Ref.~\cite{Beringer:1900zz}. We denote an amplitude for the $B$-meson, $B$, decaying to final state $f$ by $\mathcal{A}_f$.  The CP-conjugated decay is denoted by $\overline{A}_{\bar{f}}$. Since we are interested in the $s$-wave 2-body decay of the $B$, the partial decay width is given by
\begin{equation}
	\Gamma_f = \frac{1}{8\pi}\frac{p_\ast}{m_B^2}|\mathcal{A}_f|^2
\end{equation} 
where $p_\ast$ is the magnitude of the 3-momentum of one of the daughter particles. The branching ratio, $\mathcal{B}$, can then be computed from the above partial width. 

We are also interested in the CP-violating properties of the decays. For decays of charged $B$s we can define the direct CP-violation as
\begin{equation}
	A_{CP} \equiv \frac{|\overline{\mathcal{A}}_{\bar f}|^2- |\mathcal{A}_f|^2}{|\overline{\mathcal{A}}_{\bar f}|^2 + |\mathcal{A}_f|^2}.
\end{equation}
In the case of the neutral $B^0$ decay where the final state $f$ is common to both $B^0$ and $\overline{B}^0$ decays, we have to take into account
$B^0-\overline{B}^0$ mixing in defining CP-violating parameters. This occurs when $f$ is a CP eigenstate, {\em i.e.} $\bar{f} = \pm f$. The two
CP-violating parameters can be defined as~\footnote{Here we ignore the effect of CP-violation in $B^0-\overline{B}^0$ mixing which is less than 1\%.}
\begin{equation}
	C_f \equiv \frac{1-|\lambda_f|^2}{1+|\lambda_f|^2},
	\quad
	S_f \equiv \frac{2\text{Im}(\lambda_f)}{1+|\lambda_f|^2},
\end{equation}  
where 
\begin{equation}
	\lambda_f = \frac{V^\ast_{tb}V_{td}}{V_{tb}V^\ast_{td}}\frac{\overline{\mathcal{A}}_{f}}{\mathcal{A}_f}.
\end{equation}
In case of $B^0\to K^0\pi^0$ decay, neutral kaon mixing contributes an extra factor of $- V^*_{cd} V_{cs} / V_{cd} V^*_{cs}$ in the definition of $\lambda_f$.

\bibliographystyle{apsrev4-1}
\bibliography{Bdecays}

\end{document}